\documentclass[journal]{IEEEtran}
\IEEEoverridecommandlockouts

\usepackage{comment}
\usepackage{caption}
\usepackage{stfloats} 
\usepackage{booktabs}
\usepackage{diagbox}
\usepackage{tabularx}
\usepackage{multirow}  

\usepackage{varwidth}
\usepackage{amsmath,amssymb,amsfonts,graphicx}

\usepackage{hyperref}
\hypersetup{colorlinks=true,citecolor = blue}

\usepackage{graphicx,epstopdf,caption,url}   

\usepackage{arydshln} 
\usepackage{graphics}
\usepackage{color}   
\usepackage{graphicx}   
\usepackage{graphicx,epstopdf,caption,url}   

\begin{document}
%
\title{Web3: The Next Internet Revolution}

\author{
	Shicheng Wan, Hong Lin, Wensheng Gan*, Jiahui Chen, and Philip S. Yu,~\IEEEmembership{Life Fellow,~IEEE}  \\
 
        \thanks{This research was supported in part by the National Natural Science Foundation of China (Nos. 62002136 and 62272196), Natural Science Foundation of Guangdong Province (No. 2022A1515011861), Fundamental Research Funds for the Central Universities of Jinan University (No. 21622416), and the Young Scholar Program of Pazhou Lab (No. PZL2021KF0023). Corresponding author: Wensheng Gan}

        \thanks{Shicheng Wan is with the School of Business Administration, South China University of Technology, Guangzhou 510641, China.} 

        \thanks{Hong Lin and Wensheng Gan are with the College of Cyber Security, Jinan University, Guangzhou 510632, China; and also with Pazhou Lab, Guangzhou 510330, China.  (E-mail: wsgan001@gmail.com)}

        \thanks{Jiahui Chen is with the Department of Computer Science, Guangdong University of Technology, Guangzhou 510006, China.}

        \thanks{Philip S. Yu is with the University of Illinois Chicago, Chicago, USA.} 
}

\maketitle

\begin{abstract}

Since the first appearance of the World Wide Web, people more rely on the Web for their cyber social activities. The second phase of World Wide Web, named Web 2.0, has been extensively attracting worldwide people that participate in building and enjoying the virtual world. Nowadays, the next internet revolution: Web3 is going to open new opportunities for traditional social models. The decentralization property of Web3 is capable of breaking the monopoly of the internet companies. Moreover, Web3 will lead a paradigm shift from the Web as a publishing medium to a medium of interaction and participation. This change will deeply transform the relations among users and platforms, forces and relations of production, and the global economy. Therefore, it is necessary that we technically, practically, and more broadly take an overview of Web3. In this paper, we present a comprehensive survey of Web3, with a focus on current technologies, challenges, opportunities, and outlook. This article first introduces several major technologies of Web3. Then, we illustrate the type of Web3 applications in detail. Blockchain and smart contracts ensure that decentralized organizations will be less trusted and more truthful than that centralized organizations. Decentralized finance will be global, and open with financial inclusiveness for unbanked people. This paper also discusses the relationship between the Metaverse and Web3, as well as the differences and similarities between Web 3.0 and Web3. Inspired by the Maslow's hierarchy of needs theory, we further conduct a novel hierarchy of needs theory within Web3. Finally, several worthwhile future research directions of Web3 are discussed.

\end{abstract}


\IEEEpeerreviewmaketitle

\section{Introduction} \label{sec:introduction}

\begin{figure}[h]
    \centering
    \includegraphics[trim=0 0 0 0,clip,scale=0.21]{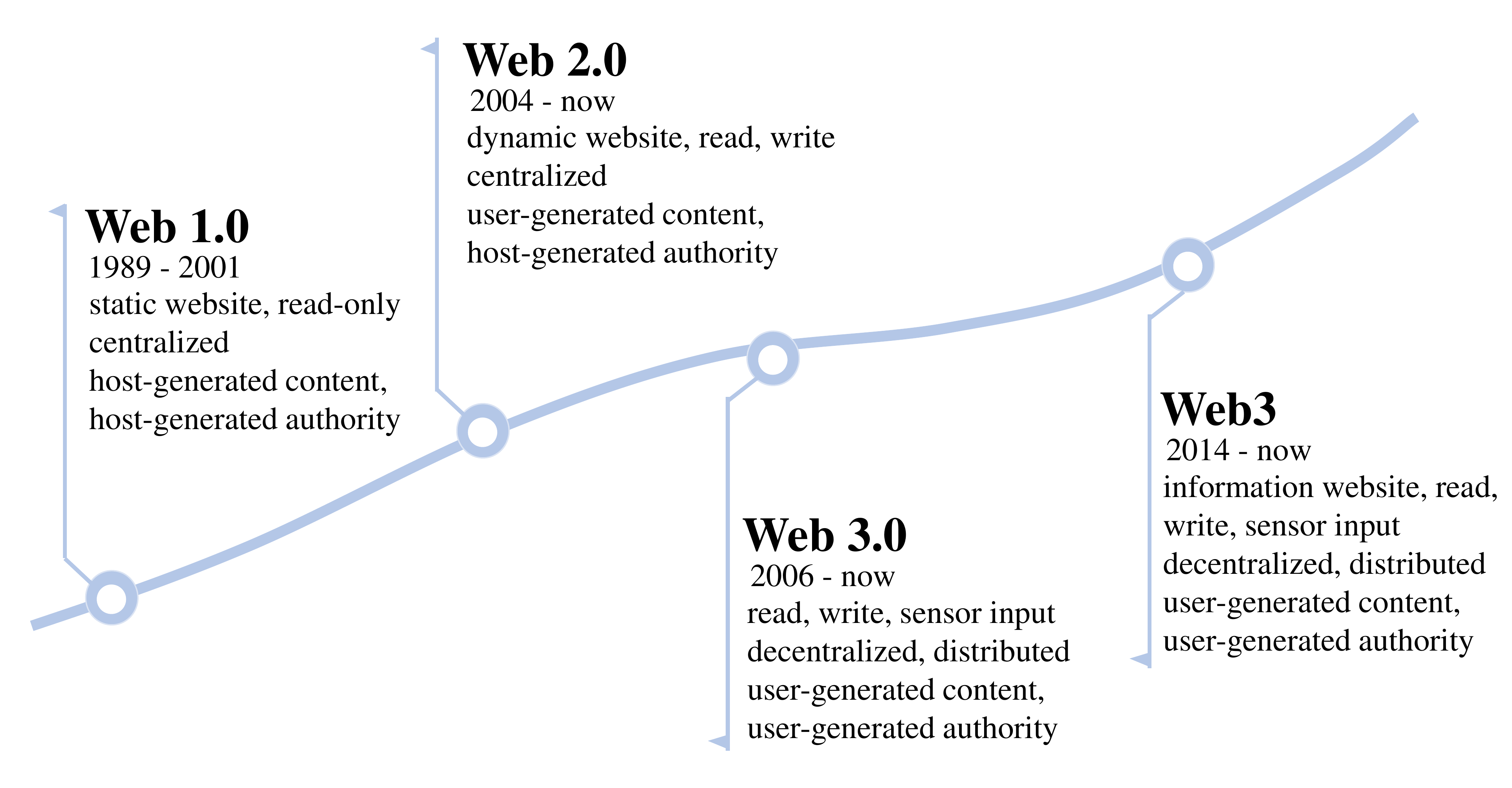}
    \caption{The development timeline of Web3.}
    \label{fig:web3timeline}
\end{figure}

As the world grapples with the COVID-19 pandemic, there are fewer opportunities for face-to-face communication. We are more dependent on the virtual world. The World Wide Web has been developed more than thirty years since the birth of Web 1.0\footnote{\url{https://en.wikipedia.org/wiki/Web1.0}}. Fig. \ref{fig:web3timeline} and Table \ref{tab:introWebs} show the history of the World Wide Web. Web 2.0\footnote{\url{https://en.wikipedia.org/wiki/Web2.0}} allows users to read and write content on the website, but the data ownership belongs to corresponding platforms. This causes a lot of monopoly companies to continuously pursue monopolizing users' data. The following Web 3.0\footnote{\url{https://en.wikipedia.org/wiki/Semantic_Web}} and Web3\footnote{\url{https://en.wikipedia.org/wiki/Web3}} are expected to break monopolies, which is the most attractive characteristic called decentralization. Web3 is the next phase of the World Wide Web and perhaps organizes society as a whole. In the Web3 era, users generate content, and the content only belongs to the user himself or herself. What we call Web3 will center around a decentralized ecosystem of technology products based on blockchain networks that are interoperable and free of traditionally trusted validators such as businesses, institutions, and government agencies.


\begin{figure}[h]
    \centering
    \includegraphics[trim=0 0 0 0,clip,scale=0.3]{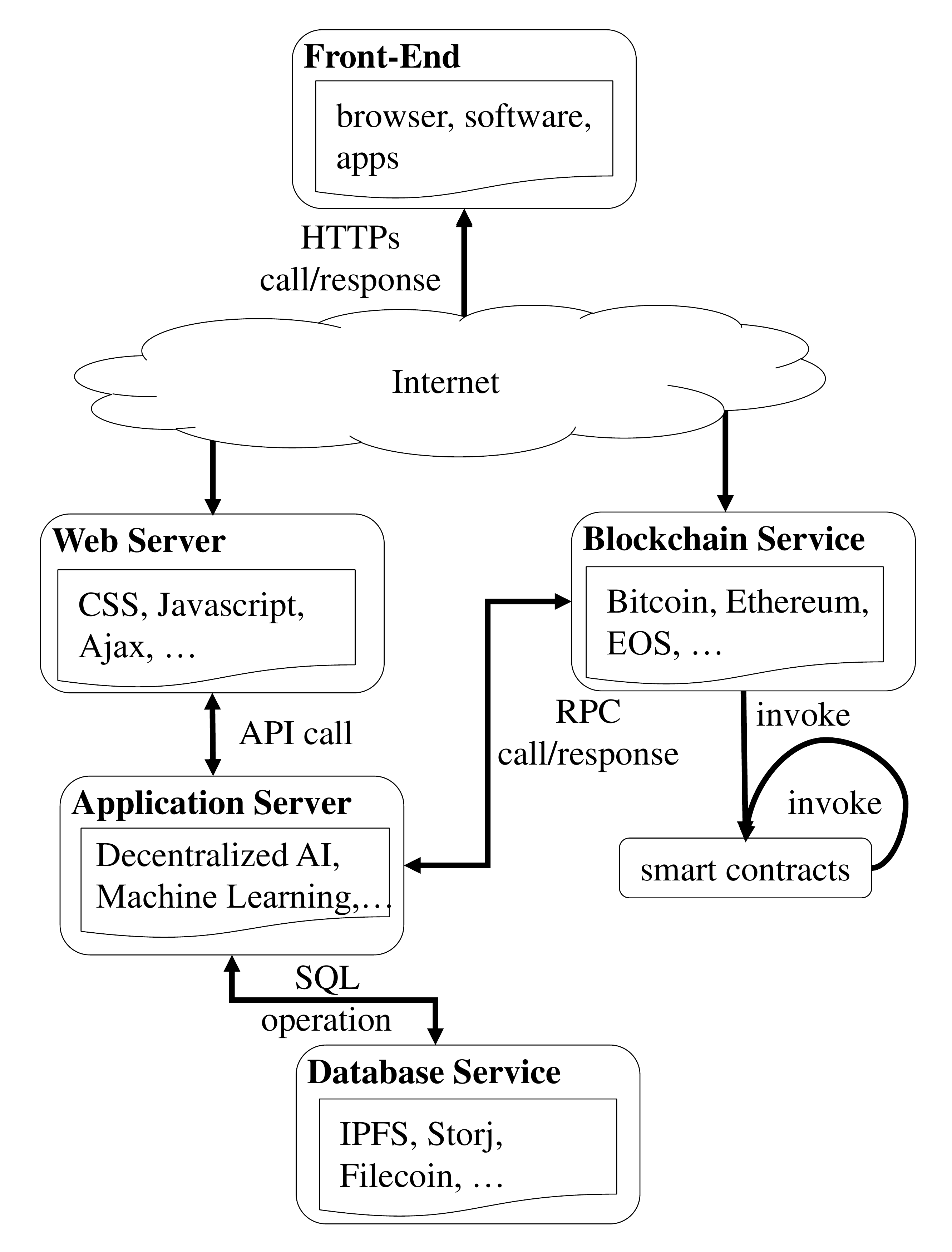}
    \caption{Web3 architecture.}
    \label{fig:web3archi}
\end{figure}

The Web3 architecture is shown in Fig. \ref{fig:web3archi}. It adopts AI technology \cite{montes2019distributed,hailemariam2020an,ekramifard2020systematic,cao2022decentralized,zolfaghari2022crypto}, machine learning \cite{li2022blockchain}, and blockchain \cite{nakamoto2008bitcoin,yazdinejad2019block,singh2020blockchain} to provide users with smart applications. This enables the intelligent creation and distribution of highly tailored content to every internet user. A digital wallet is the most important factor for each user because it offers digital identification services and user data authorization functions. After users take actions on the front end (e.g., browser and software), some request messages will be immediately sent to Web servers for calling application services. Decentralized AI, machine learning, and other big data technologies are utilized to improve the user experience and speed up the processes of systems. Furthermore, users' digital wallets will also be used to invoke corresponding smart contracts. In the meantime, the blockchain server will truthfully record users' activities on the Internet. Significantly, centralized databases will be abandoned within Web3. Distributed databases are more suitable for decentralized networks. On the one hand, it protects stored data from tampering and corruption risks. On the other hand, it provides the ability for data validation without trust. As Gavin Wood said in his blog\footnote{\url{https://gavwood.com/dappsweb3.html}}, ``the ultimate goal of Web3 is achieving less trust and more truth''.

\begin{table*}[ht]
    \centering
    \caption{A simple introduction table about Webs}
    \label{tab:introWebs}
    \begin{tabular}{| m{1cm}<{\centering} | m{2cm}<{\centering} | m{2.3cm}<{\centering} | m{1.6cm}<{\centering} | m{1.6cm}<{\centering} | m{2.5cm}<{\centering} | m{3cm}<{\centering} |}
        \hline
        \textbf{Web}  & \textbf{Period} & \textbf{Proposer} & \textbf{Generator} & \textbf{Ownership} & \textbf{Management} & \textbf{Technologies} \\ \hline
    
        Web 1.0 & 1989 - 2001 & Timothy John Berners-Lee & platform & platform & platform & data storage, data transmission \\ \hline
			
        Web 2.0 & 2004 - now  & Darcy DiNucci and Tim O'Reilly & platform, netizen & platform & platform & big data, cloud compute \\ \hline
			
        Web 3.0 & 2006 - now  & Jeffrey Zeldman & netizen, AI & netizen & netizen & peer-to-peer, RDF schema, resource description framework \\ \hline
			
        Web3 & 2014 - now  & Gavin Wood &  netizen & netizen & netizen & blockchain, smart contract, cryptocurrency \\
        \hline
    \end{tabular}
\end{table*}

The emergence of Web 3.0 and Web3 breaks the oligarchy and monopoly while creating a new productivity improvement paradigm. Therefore, it is reasonable to expect that many new methods and organizational modalities will replace the original ones. Web3 establishes a distributed network using blockchain and cryptocurrencies. The decentralized organizations are replaced by decentralized autonomous organizations (DAOs), which empower users to take ownership within the network. Furthermore, the combination of DAO and Web3 offers a possible way for addressing the existing issues. For example, decentralized science (DeSci) mainly aims to help researchers eliminate reliance on profit-seeking intermediaries. The decentralized social blockchain (DeSo) distinctly differs from the centralized social applications and solves the profit distribution issues currently faced. The DeSo developers have fundamentally designed a blockchain architecture that is suitable for massive social activities. The revenue generated from user creation will be recorded in users' digital wallets. To be noticed, the DeSo does not provide decentralized finance.

\textbf{Research gap}: For a long time, people have been subjected to monopolies in the Web 2.0 era. The ownership of data is visible with the help of Web3 which gives control back to users. A wonderful future world is waving at us, but there still exist some research gaps that need to be filled. Firstly, Web3 sounds like a buzzword, as only a few people can accurately describe what real Web3 is. Secondly, though the current literature \cite{rudman2016defining,keizer2021case,korpal2022decentralization,chohan2022web,wang2022exploring,nabben2023web3,murray2023promise} had already taken review on Web3 domain, most of them often confused Web 3.0 with Web3. Thirdly, since the necessary conditions for the sustainable growth of Web3 have yet to be mature, such decentralized applications have been used by a few people in Web3's infancy. Besides, most of us believe that Web3 will be a huge wave that deeply changes our lives. Indeed, discussing Web3's challenges, risks, opportunities, and future directions is critical.

\textbf{Contributions}: To fill this gap, this article aims to conduct a systematic literature review of Web3. This article contributes to providing helpful guidance and support for Web3 future research and industrial applications. The contribution of the article is fourfold. 

\begin{itemize}
    \item This is the first article that takes an up-to-date overview of the development and challenges of Web3. We discuss the advantages and disadvantages of Web3 technologies. The introduction investigates the differences among the four stages of the Web and gives a Web3-based architecture.
    
    \item We illustrate the main technologies and applications within Web3. We briefly describe their roles and functions and how they have an impact on our lives.

    \item Due to the definitions of Web 3.0 and Web3 being confusing in most current literature, we conduct a discussion about the differences and similarities between them. Furthermore, we also focus on introducing some challenges and issues that Web3 may bring.

    \item Inspired by Maslow's hierarchy of needs theory, we attempt to propose a novel hierarchy of needs theory within Web3. This article also examines some current issues, such as privacy, security, law, and government, and then suggests some directions for future research.
\end{itemize}

\textbf{Organization}: This article discusses Web3 in detail. Section \ref{sec:technologies} introduces some major technologies within the Web3 architecture. Then, we discuss some different types of Web3 applications in Section \ref{sec:applications}. In Section \ref{sec:concerns}, we discuss some differences and similarities between Web3 and Web 3.0. We also give a deep discussion of Web3 in the rest of the subsections. Sections \ref{sec:challenges} and \ref{sec:directions} introduce some challenges and future research directions, respectively. At last, we simply conclude this article in Section \ref{sec:conclusion}.

\section{Web3 Main Technologies} \label{sec:technologies}

\begin{figure}[h]
    \centering
    \includegraphics[trim=0 0 0 0,clip,scale=0.25]{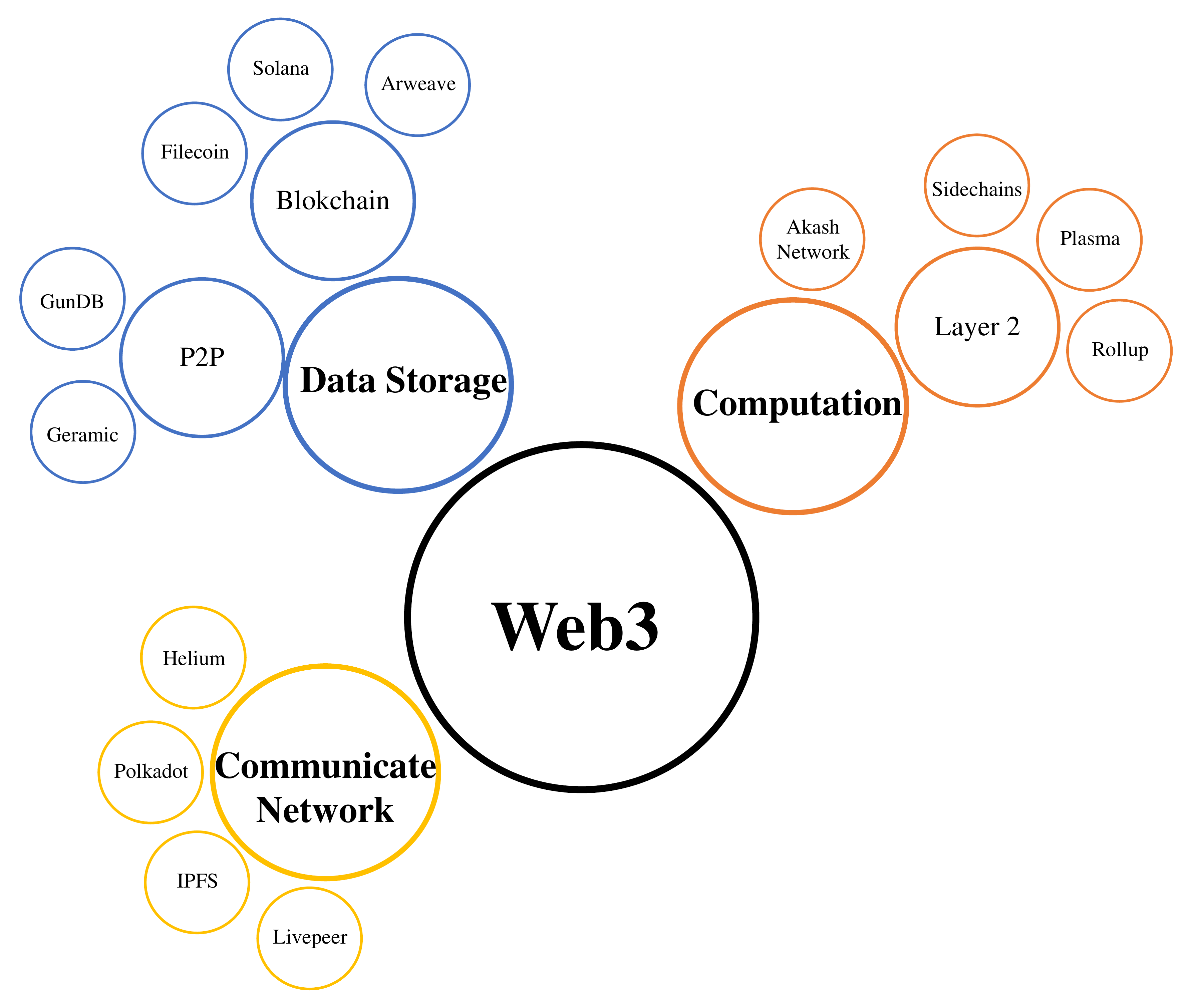}
    \caption{Several main technologies of Web3.}
    \label{fig:web3technologies}
\end{figure}

We provide a list of common abbreviations in Table \ref{tab:abbreList} to simplify the following discussion. As shown in Fig. \ref{fig:web3technologies}, some main technologies of Web3 can be roughly divided into three parts: data storage, communication networks, and computing. These technologies work together to ensure that the decentralization property can be implemented in new applications.

\begin{table}[h]
  \small
    \centering
    \caption{A list of common abbreviations}
    \label{tab:abbreList}
    \begin{tabular}{m{0.8cm}<{\centering}  m{4.5cm}<{\centering}  m{2cm}<{\centering}}
    \hline
    \textbf{Ref.} & \textbf{Full name} & \textbf{Abbreviation}  \\ \hline
    \cite{raman2019challenges} & Decentralized Web & DWeb  \\ \hline
    \cite{wood2014ethereum} & The third stage of World Wide Web (blockchain-based) & Web3  \\ \hline
    \cite{berners1998semantic} & The third stage of World Wide Web (semantic-based) & Web 3.0  \\ \hline
    \cite{benet2014ipfs} & InterPlanetary File System & IPFS \\ \hline
    \cite{wang2019decentralized} & Decentralized Autonomous Organization & DAO  \\ \hline	
    \cite{wu2021first} & Decentralized Application & Dapp  \\ \hline
    \cite{schar2021decentralized} & Decentralized Finance & DeFi  \\ \hline
    \cite{diaz2022governance} & Regenerative Finance & ReFi \\ \hline
    \cite{ding2022desci} & Decentralized Science & DeSci  \\ \hline
    \cite{li2019incentivized} & Decentralized Social Blockchain & DeSo  \\ \hline
    \cite{wang2021non} & Non-Fungible Token & NFT  \\ \hline
    \cite{karandikar2021blockchain} & Fungible Token & FT  \\ \hline
    \cite{nakamoto2008bitcoin} & Blockchain Technology & BT  \\ \hline	
    \cite{weiser1999computer} & Internet of Thing & IoT  \\ \hline
    \cite{montes2019distributed} & Decentralized Artificial Intelligence & DAI \\ \hline
    \cite{zhao2019secure} & Secure Multi-Party Computation & SMPC \\ \hline
    \cite{zhu2022blockchain} & Trustworthy Federated Learning & TFL \\ \hline
    \cite{sabt2015trusted} & Trusted Execution Environment & TEE \\ \hline
    \end{tabular}
\end{table}

\subsection{Data Storage}

The research around the area of distributed databases, such as P2P databases, has been developing for a long time \cite{ozturk2015warehousing,huh2019xr}. A popular distributed ledger of transaction databases, i.e., blockchain \cite{nakamoto2008bitcoin}, is simply a natural infrastructure of Web3. The blockchain is famous for its immutable and append-only structure. This property ensures that completed transactions cannot be erased or forged, which protects the system from corruption. Blockchains will take several transactions as a ``block'', and then the next ``block'' will connect the previous one to form a ``chain''. The new block also contains the cryptographic hash of the previous block according to a Merkle Tree \cite{merkle1987digital}. Therefore, if someone (e.g., attractors) tries to modify data in a block, the linked block will not match the cryptographic hash of the changed block anymore. Otherwise, this user has to change all blocks on the chain at the same time (it is almost equivalent to creating a new chain). Such is the ingenuity of the blockchain structure. Meanwhile, the immutable property also helps to solve one of the core problems of Web3: data exchange between different websites. In the Web 2.0 area, different websites can freely read and write the data of the same user. However, the recorded and changed data cannot be easily shared with other platforms, because data is a cash cow. This causes users to have to register accounts at different websites again and again. It not only raises the risk of private information leakage but also burdens users with extensive data management. Once Web3 is built on the blockchain, according to the immutable nature of the blockchain, a unique digital wallet represents users' ID cards and bank accounts in the virtual world. The digital wallet can function as a privacy key management tool to identify who you are, and the private key is something only the user himself or herself knows at the same time. On the other hand, with the help of a digital wallet, financial activities in the virtual world will be more like those in the real world, to a certain degree.

\subsection{Communicate Network}

A communication network is a pattern that aims to effectively organize communication information. It is also the established system where messages can be sent to one or more directions within the organization based on the requirements. The InterPlanetary File System (IPFS) \cite{benet2014ipfs} has been widely adopted in Web3, which is expected to replace the traditional Hyper Text Transfer (HTTP) \cite{fielding1999hypertext}. It helps people search the internet through a content-addressable mechanism instead of an IP-addressable mechanism. For example, if you want to watch or download a movie (such as Titanic, Avatar, or Harry Potter), the browser first has to know where the movie is on the Internet (i.e., the domain name or IP address). However, if the servers are unfortunately closed, you will get ``404 Page Not Found'' error from browsers. That means you cannot watch or download the movie. Others may have downloaded the movie before and still have it, but it seems you have no way of connecting with that them. However, the situation is totally different within IPFS. Each file in IPFS will have a unique hash value based on file content. Hence, you just need to tell your browser what you want, and then the browser will search for the target hash value from IPFS. Additionally, there is no blockchain that is capable of supporting all kinds of functions and being applied in distinct situations. The Bitcoin blockchain can deal with transactions among accounts very well and is more likely a value-exchange medium \cite{nakamoto2008bitcoin}. The Ethereum blockchain is designed to act as a platform and promote the development of smart contract technology. Polkadot can link several blockchains into an integrated network. In Web3, collaboration among different blockchains will be more and more important. This means that blockchains are able to offer better services while improving their efficiency and security by reducing unnecessary code. In short, building rich, information-sharing, and cross-platform responsive web applications is also needed with the help of cross-blockchain collaboration \cite{kazemi2021towards}.

\subsection{Computation}

The generated time cost of a block in Bitcoin is around ten minutes, while Ethereum processes 10 to 15 transactions per second. This case reveals that the processing speed of Web3 will be a visible barrier. As a result, numerous academics and developers proposed a variety of solutions. Let's take Ethereum as an example. On the one hand, the blockchain demands that all nodes take part in recording transactions. On the other hand, the massive trading volume on blockchain causes communication crowding. The above issues always increase latency and potentially result in a worse user experience. Therefore, additional chains are proposed, where the Layer 2 (additional) chains \cite{sguanci2021layer} are added to the main chains (that is, the Layer 1 chains, e.g., Bitcoin and Ethereum), and process new transactions faster while reducing the load on the main chains in the meantime. Though sidechains \cite{parizi2019interg} function as auxiliary networks to a Layer 1 chain, the sidechains have their own consensus approaches and therefore run independently with main chains. If data from sidechains is concealed, forged, or tampered with, the dealing results that the main chains receive are obviously wrong. After that, the plasma chains \cite{poon2017plasma} are likely a tree structure, which compresses the representation of transaction results and then commits it at a fixed interval. The difference between plasma chains and sidechains is that plasma will check the correctness of transaction results before sending them to the main chains. If an error occurs, users are able to safely withdraw from plasma chains. What's more, rollup chains \cite{gramoli2015rollup} compress a certain amount of transaction data on a block at some intervals. The main chains not only accept the transaction results but also receive information about the transaction. Up until now, rollup has been the topic of Layer 2 solutions because of its high throughput and low cost. Furthermore, the optimistic rollup and the ZK rollup are both practice solutions with rollup chains. ZK rollup adopts zero-knowledge proof \cite{feige1988zero} to make sure confirmation and a tamper-proof record of transactions. It reduces the validity proof cost but does not effectively save computation. Optimistic rollup is a solution for extending universal smart contracts, but it suffers from the same flaw as plasma chains.

\section{Types of Applications of Web3 Developers} \label{sec:applications}

\subsection{Web3}

\begin{figure}[h]
    \centering
    \includegraphics[trim=0 0 0 0,clip,scale=0.23]{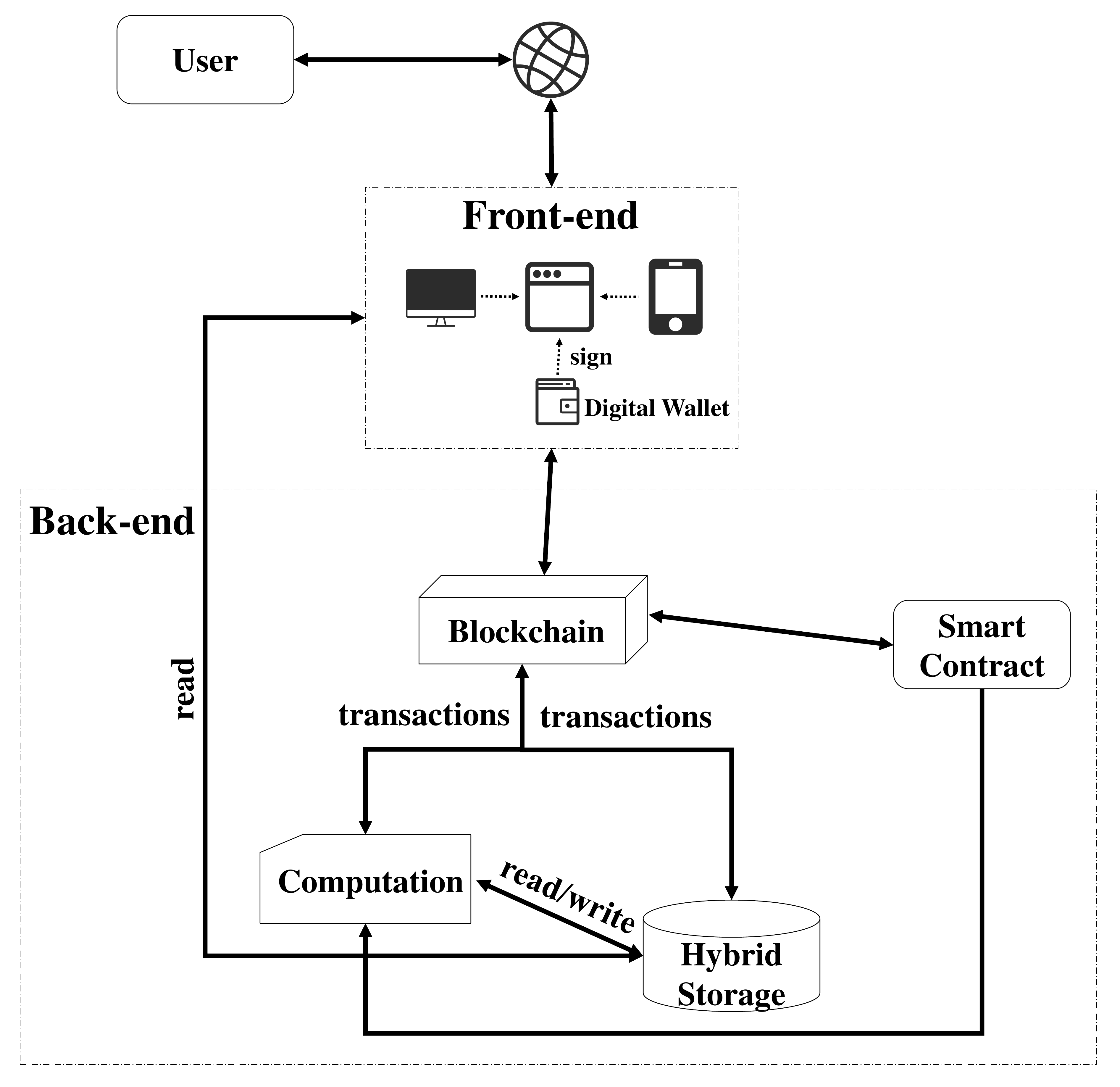}
    \caption{A simple instance of Web3.}
    \label{fig:web3case}
\end{figure}

While Web 2.0 was a front-end revolution, Web3 is an in-depth back-end revolution. Web3 embraces a set of protocols based on blockchain, which intends to reinvent how to return data ownership to users and let everyone equally participate in it. Fig. \ref{fig:web3case} reveals a simple instance of a user case within Web3, and we provide a detailed introduction to describe how the Web3 system works in this subsection.

Tom recently plans to buy a new smartphone, and he also wants to sell the old one. He opens a Web3 software or website (like decentralized eBay or Letgo) and then publishes a smart contract. The contract records information about the sale of his phone, such as the selling price of \$200, blue color, complete appearance, usage time, warranty date, and ownership. Jim learns from a decentralized Web3 compatible service that Tom has posted a selling smart contract. Jim wants to buy the phone and clicks on ``buy'' button. The smart contract services first check that the selling phone indeed belongs to Tom, and then make sure Jim has enough tokens (\$200) in his digital wallet. If the results on both sides are true, the smart contract is automatically enforced. The ownership of the phone will be re-registered as Jim instead of Tom, and the balance of Tom's digital wallet increases \$200. At last, Jim receives Tom's phone, and then this transaction is done. Maybe someone will ask: ``if Tom refuses to give the phone to Jim, what will happen?'' Firstly, the smart contracts already change the ownership of the phone from Tom to Jim. Secondly, the phone cannot offer an unlock service for Tom without Jim's authorization. Thirdly, the blockchain network can easily verify the whole transaction process between them. Therefore, it makes no sense for Jim to keep this phone.

\subsection{Blockchain}

Herein, we further explore Web3 from the perspective of blockchain. A blockchain runs with a lot of automatic protocols. These protocols offer a variety of different web services (e.g., storage, transaction, and identity). There are also some trigger protocols that replace unnecessary and inefficient intermediaries in Web 2.0. A new song, for example, begins with creation and ends with a publication, which is frequently completed through collaboration. A third-party company is usually in charge of things like the release, rate payment, and profit distribution among collaborators. There are so many participants involved that copyright or profit disputes often happen. Nevertheless, things will be simple with the help of blockchain technology. Once fans confirm the purchase of a song, a certain number of tokens will immediately be sent from their digital wallets to the creator's digital wallet. The entire process is automatically executed by several triggered protocols. If there are several co-creators, the blockchain faithfully records the contribution of each one. The information is open and transparent, and no one can have exclusive access to it. In the Web3 era, there will be a lot of talented individual creators. There is no need to share their revenue with intermediaries, and this is good news for both creators and supporters. The immutable property will make sure a creator enjoys rights over their works throughout its lifetime.

\subsection{Smart Contract}

The smart contract \cite{singh2020blockchain} is a kind of digitalized contract in the virtual world. It will automatically execute if some conditions are met. Smart contracts were proposed in the early years, but due to the trust issue, they have not been widely used in recent decades. After all, no one will unreservedly believe a stranger will keep their contracts without a reliable guarantee. Fortunately, blockchain technology provides a trusted virtual environment and then promotes the application of smart contract technology. Smart contracts allow developers to build new applications on top of blockchains. For instance, smart contracts can be written to create systems such as authentication, financial transactions, supply-chain management, power distribution, and reputation systems. What's more, the importance of smart contracts on distinct blockchains is different. Bitcoin \cite{nakamoto2008bitcoin} can be seen as a naive smart contract that deals with peer-to-peer transactions. Ethereum \cite{wood2014ethereum}, on the other hand, is built as a decentralized computation network. As a matter of fact, Bitcoin is more likely a neutral bank. It only offers fast and convenient payout and deposit services and does not care how or why you do anything. The smart contracts in Bitcoin can be seen as a complete set of operation-locking protocols. It targets a single mission: protecting your tokens and ensuring tokens are sent to the right digital wallet. On the other hand, Ethereum regards any contract as essentially a series of transactions. The state of the last transaction is related to the state of the pre-transactions. So, it provides a general mechanism for tracking and updating historical states, and users can customize the corresponding logic and conditions. In a word, smart contracts are verified by the peer-to-peer network. This feature avoids the involvement of a third party. Furthermore, automating smart contracts frees massive amounts of manual services.

\subsection{Decentralized Finance}

Decentralized finance (DeFi) \cite{schar2021decentralized} refers to decentralized applications of finance, including exchange, loans, and investment. DeFi is a global, open-source, and accessible financial and technological infrastructure for all citizens of the internet. Blockchain technology enables nearly instantaneous and peer-to-peer digital transactions. Before the birth of Bitcoin, centralized intermediaries, such as banks and credit card institutions, could deal with almost all digital payment activities. The perfection of bank infrastructure has a great impact on financial liquidity. Cryptocurrencies are like money in Web3 \cite{gely2016cryp}. Cryptocurrencies are digital currencies underpinned by cryptographic systems, where cryptography ensures the security of online payments without the use of centralized third-party intermediaries.

Aside from the features of DeFi we've already talked about, DeFi services also use other Web3 features, such as open governance and low fees. Financial inclusion, which provides affordable and user-friendly financial products and services, is vital for the economic cycle. Nowadays, if you are living in an emerging country, perhaps activities such as payments, savings, credit, and insurance will not be easy without financial and technological experience. Today, DeFi services are being experienced in real life by more than one million people. Through user-friendly operation and low-cost automated financial services that occur at scale, DeFi services could potentially evolve and lead to the broader global financial inclusiveness of billions of unbanked people worldwide in the future.

\subsection{Non-Fungible Token} 

Non-fungible token (NFT) \cite{wang2021non} refers to a special kind of digital token. It is the opposite side of a fungible token (FT) \cite{karandikar2021blockchain} like bitcoin. In practice, FT represents currency, shareholding, and other quantitative differences. NFT represents a collection, an ID account, an IP address, and other unique assertions. An FT can be divided into several equivalent assets, and each asset can be freely further divided or combined. However, an NFT-like digital collection cannot be divided and is usually used to represent specific and unique things. In other words, a bitcoin is equal to a bitcoin, but a digital collection cannot be regarded as another digital collection.

Compared to FT, the liquidity of NFT is obviously lower. However, NFT also plays a vital role in peer-to-peer transactions in Web3. Firstly, NFT is friendly to individual creators. In the Web 2.0 era, creators have to upload their digital products to online platforms (e.g., Instagram, YouTube, TikTok, and Spotify), and then these centralized platforms promote products and earn revenue in different ways. Creators and platforms share profits. In contrast, digital creators can directly earn profits from their fans via NFTs. Because NFTs are public on the blockchain, buyers rarely purchase fake products and get a clearer picture of the creation process. In addition, NFTs are also able to contribute as much as possible to copyright registration, and identification recognition in Web3.

\subsection{Decentralized autonomous organization}

Decentralized autonomous organization (DAO) \cite{wang2019decentralized} is a blockchain-based and internet-native organization model that encodes management and operating rules according to a series of smart contracts. DAO is also a staple within Web3, which discards centralized control or third-party participation elements. There are three basic elements that a DAO owns:

\begin{itemize}
    \item What the mission of the DAO is? A clear mission is the first step for consensus-building among members. Its common goal can quickly gather a number of strangers to work together.
    
    \item How to maintain consensus in a DAO? A well-developed organization depends on a complete rule-based system, including governance, incentives, voting, and punishment. These rules are encoded by smart contracts and deployed on the blockchain, which prevents them from being changed.
    
    \item What can an incentive involve? An appropriate incentive mechanism ensures that members contribute to the sustainable development of the DAO. In some cases, incentives help discover talented individuals from the masses and encourage them to play greater roles.
\end{itemize}

Thereafter, we briefly introduce how DAO works: Firstly, DAO founders have to create a new kind of cryptocurrency\footnote{That cryptocurrency is not the major cryptocurrency (e.g., Bitcoin, Ethereum, and Binance coin) in most times.}. Secondly, founders release cryptocurrency to stakeholders (DAO members). Finally, his or her voting power and influence are determined by the amount of cryptocurrency he or she owns. Additionally, cryptocurrency can be bought and sold at will between DAO members.

\subsection{Metaverse}

The Metaverse \cite{sun2022metaverse,lin2022metaverse,chen2022metaverse} is a completely new social framework that will eventually affect every aspect of our lives. The connection between Metaverse and Web3 appears to be a mapping between the productive force and the production relation. Two pivotal elements of the Metaverse (i.e., digital identity and the economic system) are closely related to the maturity of Web3 technologies. Web3 ensures that the production relationship is always people-oriented, democratic, open, and global in nature. In the Metaverse era, just as electrical technology played a role in the Second Industrial Revolution, Web3 owns the underlying web-service technology and economic support. It will promote the growth of productive forces. Additionally, before we discuss the relation between Web3 and Metaverse, it should be noted that the development of Metaverse is not dependent on Web3, and vice versa, because Web 3.0 can also play the same function \cite{webGan2023}.

In practice, both Web3 and Metaverse are capable of impacting the future virtual world to a certain degree. Virtual reality and artificial intelligence technologies meet the requirements for the front-end of the Metaverse. The innovation of blockchain and big data technologies within Web3 offers powerful technical support for the growth of the back-end of Metaverse. Metaverse illustrates how productivity has changed in the Internet age, from computers to mobile phones to VR and AR devices. By contrast, Web3 is more likely an innovation of the regime. It makes a difference in the production relationships that return data ownership to users. The combination of Web3 and Metaverse can be thought as a novel decentralized Metaverse. It creates a new economic model in the Metaverse, i.e., the creator economy. Web3 provides a new method and novel paradigm for copyright, circulation, and value discovery of digital products. This case will deeply invoke the passion of creators, which leads to a better experience for others. Web3 will encourage more and more users to participate in the decentralized Metaverse, which will put all bribes back into Web3 and Metaverse in the end.

\section{The Current Web3 Concerns} \label{sec:concerns}

Recently, Stack Overflow completed a 595 developer questionnaire survey, and the result post title is ``New data: Do developers think Web3 will build a better internet?'' \cite{gibson2022new}. The result shows that most people surveyed had already heard or learned about Web3, but only a quarter thought that Web3 was the future of the internet. The popularity of blockchain and Web3 is undeniable, but the question is, will this all stand the test of time?

\subsection{Web3 vs Web 3.0}

\textbf{Difference between Web3 and Web 3.0}. Though both Web3 and Web 3.0  \cite{webGan2023} are familiar names, there are several remarkable differences between their technologies and concepts.

\begin{itemize}
    \item Web 3.0 \cite{webGan2023}, also known as the semantic web, focuses more on efficiency and intelligence by reusing and linking data across websites. What's more, Web3 puts a strong focus on security and returns data ownership back to users.

    \item Web 3.0 adopts the solid pod\footnote{\url{https://solidproject.org/}} to store all user data, where the third party has to request access authority for reading user data. Each user in the digital ecosystem will be assigned a unique WebID as their identity via a solid pod. Web3 utilizes a cryptocurrency digital wallet, which mainly depends on users' privacy keys, to identify them.

    \item In addition, both Web3 and Web 3.0 technologies provide data security. The major difference is the methods adopted to reach this purpose. Web3 is blockchain-based, while in Web 3.0, new variations of traditional data transmission and empowerment technologies like Resource Description Framework Schema, Web Ontology Language, SPARQL Protocol, and RDF Query Language are utilized.

    \item Since blockchain is a distributed database, data in Web3 is difficult to modify or delete; however, data in Web 3.0 may be changed effortlessly because data may be stored in a centralized database. For instance, a solid pod may use an unsafe storage approach like a centralized platform, while crypto wallets on the blockchain can only be accessed by users' private keys.
\end{itemize}

\textbf{Similarities between Web3 and Web 3.0}. Though Web3 and Web 3.0 have many differences \cite{webGan2023}, they share a common purpose. Both Web3 and Web 3.0 aim to give back data ownership to users and then build a really free and fair web. In terms of data storage, Web3 and Web 3.0 are both likely to adopt a distributed model to enable decentralization. Whatever the next generation of the Web is, both Web3 and Web 3.0 are ideal virtual worlds for which scholars and developers strive. Democracy, harmony, freedom, equality, and justice will be further incorporated into our internet lives. Furthermore, Web3 and Web 3.0 are still in their early stages. Despite the publication of massive business products with Web3 and Web 3.0, they have yet to walk a long growth path.

\subsection{Data Storage in Web3}

As users' awareness of data ownership grows, the disadvantages of Web 2.0's data storage model (e.g., centralized storage) are gradually exposed. Commercial companies require users to trust them in order to provide storage services. While users enjoy convenience, they also lose ownership of data, because operators may change user data intentionally or unintentionally at any time. Web3 will use a new storage model, namely decentralized storage. Nevertheless, even though Web3 runs on blockchain, it cannot be achieved solely on blockchain. There are some problems in using the blockchain to store all data \cite{akash2022how}. On the one hand, the blockchain's transaction speed is far from adequate for dealing with massive amounts of data. On the other hand, storing massive data violates the design of the blockchain. The original idea was to store only a small amount of representative data for sharing and verification among nodes. For example, an NFT on the blockchain doesn't actually store the actual item it represents, but just a pointer to data that is not on-chain. Therefore, if these data are stored in a centralized manner, once they are changed, the NFT will become a null pointer. Obviously, this cannot guarantee the lifetime and unique characteristics that NFT claims. In addition to ensuring users' data ownership, decentralized storage is more cost-effective than centralized storage \cite{chris2022storj}.

At present, there are already some decentralized storage solutions. The solution of Storj\footnote{\url{https://www.storj.io/}} is to divide the encrypted data files uploaded by users into several blocks, and then distribute these blocks to each storage node. These nodes come from those storage space contributors who benefit from the incentive mechanism. Only when a certain number of blocks are met can the corresponding file be retrieved. This can ensure that the complete file cannot be obtained when a few storage nodes are hacked \cite{moe2022web3}. IPFS is a hypermedia protocol that is a counterpart to the HTTP protocol. It has the characteristics of peer-to-peer. The content loads no longer come from the central server but are retrieved from several storage nodes. In addition, the addressing is content-based, rather than traditional location-based (such as domain name and IP). Filecoin\footnote{\url{https://filecoin.io/}} is a decentralized network for storing files built on IPFS. It has a built-in incentive mechanism and issues a native cryptocurrency to facilitate transaction efficiency. Also, there are other solutions (such as SIA\footnote{\url{https://sia.tech/}} and SWARM\footnote{\url{https://www.ethswarm.org/}}). In short, what they all have in common is that the complete data file is chunked and stored with the help of multiple nodes in the network. The storage nodes only have the information of the chunks and not the entirety of the file, thus ensuring that only the file owner can access it \cite{moe2022web3}.

\subsection{Privacy and Personal Data in Web3}

In the Web3 world, some people who dislike the poor underlying technologies will do some nesting based on the original infrastructure. However, the problems brought about by these ``stacked houses'' may exceed the harm caused by outdated planning and even become a worse fusion of the shortcomings between Web 2.0 and Web3. For example, after NFT becomes popular, fractional NFT (F-NFT) occurs, which uses the idea of FT to dismantle NFTs. The Centralized DeFi (CeDeFi) has been proposed to modify the traditional cryptocurrency model and is expected to hold costs down. Several subsequent thoughts are conducted using the simple examples of the above scenario applications. Firstly, the fusion of centralization and decentralization ideas in products is perfect, but it may be a castle in the air. On the one hand, this kind of fusion approach is more complicated than utilizing either centralization or decentralization alone. On the other hand, because of the lack of supervision among CeDeFi, many frauds have happened in the last few years. Second, due to the astronomical value of personal data on decentralized platforms, this type of fusion application or software still exists, with the risk of intentional information leakage. Thirdly, compromise is a kind of wisdom on security issues. It suggests striking a balance between the costs of protection and attack. Considering both centralization and decentralization properties at the same time undoubtedly adds to the development burden.

\subsection{What Problem Does Web3 Solve?}

Identity authentication at the Web 2.0 stage is a necessary but troublesome thing for users. Due to the inconsistent authentication of different platforms, we often forget passwords, and it is not easy to recall email addresses and security questions when resetting passwords. Web3 will solve these problems with the combination of a digital wallet and digital signing, providing a common way to represent the ownership of data, assets, and certificates. Web3's wallet is a fixed address that is bound to the user's identity and stores the user's asset information. Users prove their ownership of the wallet by providing a digital signature \cite{yazdinejad2020dec}. Obviously, they still need to spend energy to remember the private key, but since the wallet is directly linked to the asset, there will be great motivation to drive them to protect their private key \cite{bambacht2022web3}. If users own any NFTs or other tokens, they can connect their wallets to different websites and applications, view and manipulate these assets in different ways, and it will all be unimpeded. A number of products already exist and implement the functionality of Web3 wallets, including MetaMask, WalletConnect, Web3Auth, and Forform, which serve as apps or browser extensions.

In addition to solving the issue of authentication, Web3 also has a set of advanced solutions to address the issue of community governance. The current common centralized system of governance is based on discussion and voting, with venues usually being social media, online forums, meetings, etc. Web3 will enable a new model that addresses the shortcomings of traditional governance models through blockchain. Governance on the chain has many advantages, including a transparent process and precise accountability \cite{potts2022exchange}. One solution is the issuance of governance tokens, which have become the fundamental decision-making mechanism for DAOs. They are a type of cryptocurrency with immutable ownership and transparent distribution, which makes them ideal for decentralized decision-making \cite{marcel2022what}. By voting, token holders can have an impact on the development of new products or features.

\section{Challenges and Issues} \label{sec:challenges}

Web3 is still in its early stages. It also provides a novel version of the web that lets users have more control over their internet data and profit distribution. Furthermore, the rise in risks and challenges is accompanied by the expansion of Web3. Though some Web3 products have come out with solutions for centralization and financial activity issues, they may cause some new variations on these problems.

\textbf{Several risks and problems within Web3.} The blockchain is one of the core components of Web3, which means that the classic ``Impossible Triangle'' problem of blockchain technology is inevitable. The problem refers to the fact that decentralization, scalability, and security cannot be compatible at the same time \cite{qin2018overview}. For instance, some Web3 products increase the number of users by improving scalability and reducing the generation time of blocks to accelerate computation speed and cost \cite{wang2022exploring}. Though the approaches described above aim to improve the user experience, they increase the likelihood of making the blockchain vulnerable to hacking. Hence, researchers in Web3 should consider how to make compromises and trade-offs among these three vital properties. Another severe issue that is a barrier to the development of Web3 is financial fraud (which includes cryptocurrency and NFT). The most common way is called ``exit scam\footnote{\url{https://en.wikipedia.org/wiki/Exit_scam}}''. Encryption developers use various false propaganda to attract early investors to participate in a project, and then abandon the project or take away users' investment in a short time. Since cryptocurrency can be launched on a decentralized exchange without any audit, ``exit scam'' within cryptocurrencies gradually becomes a new kind of low-cost crime \cite{sheridan2022web3}. In addition, it seems that the usage scenario for NFT is restrictive. Considering a simple instance of the asset transformation case, Tom plays two shooting games (such as CSGO and PUBG) in the meantime. Does it make sense to take a weapon from CSGO to PUBG? Maybe someone will give some examples, such as the exchange of skins or characters, but keeping the ``game balance'' is not easy. Thus, bringing Web3 technologies into deeper consideration is challenging but needed.

\begin{figure}[h]
    \centering
    \includegraphics[trim=0 0 0 0,clip,scale=0.2]{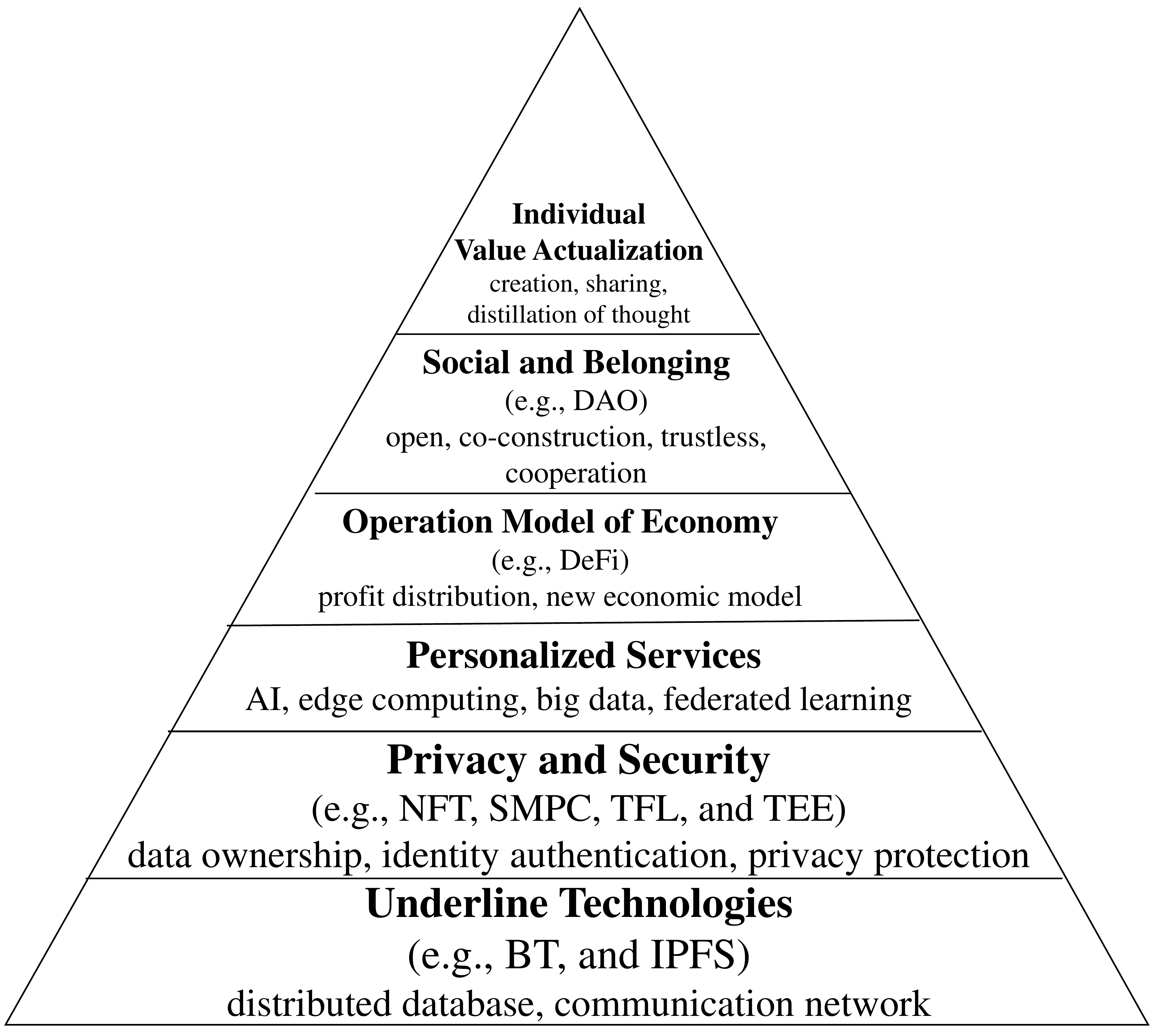}
    \caption{The hierarchy of needs within Web3.}
    \label{fig:web3hierarchical}
\end{figure}

\textbf{The new hierarchy of needs within Web3.} Inspired by the Maslow's hierarchy of needs\footnote{\url{ https://en.wikipedia.org/wiki/Maslow\%27s_hierarchy_of_needs }}, we propose a novel hierarchy of needs within Web3, as shown in Fig. \ref{fig:web3hierarchical}. Firstly, \textbf{underline technologies} (e.g., distributed database and communication network) provide basic frameworks and supports for human needs on the Web. The distributed database can ensure that data ownership belongs to users themselves to a great extent, while a redundant backup strategy protects users' data from corruption and falsification. Secondly, \textbf{privacy and security} not only ensure that Web3 applications hold on the decentralization property, but also intuitively affect the interests of users. Therefore, the trusted execution environment (TEE) \cite{sabt2015trusted, yazdinejad2020cost} functions between software and hardware, which provides a low extra cost and medium protection framework for device security. TEE allows developers to create secure applications because it provides rich user interfaces and external connection capabilities. TEE can safely get users' tokens (like NTF) and make identity authentication according to trusted user interfaces. Nevertheless, the pursuit of security protection may end up invading personal privacy. Therefore, trustworthy federated learning (TFL) \cite{zhu2022blockchain} and secure multiparty computation (SMPC) \cite{zhao2019secure} ensure that data leakage does not occur while Web3 products authenticate identity.

Against the background of a rebalancing towards privacy and security considerations, it is possible that humans achieve higher-level needs. The third layer is \textbf{personalized services}. Based on big data and artificial intelligence techniques, individualized service lets everyone obtain a series of customized solutions including education, training, and recreation \cite{lin2022metaverse}. After obtaining a good education, people will also seek a decentralized and interest-driven global economic system. That is the fourth layer of the hierarchy of needs within Web3 (\textbf{operation model of economy}). The profit distribution will be fairer and more transparent. Last but not least, as a famous saying goes, ``The economic foundation absolutely determines the superstructure.'' The novel economic model will result in new relations among people, organizations, and people themselves. That's why the fifth layer (\textbf{social and belonging}) is required. The decentralized economic model will promote a future social environment that is more open, trustless, and cooperative. Finally, the top need (\textbf{individual value actualization}) is the ultimate goal of human development. As a quotation says, ``What a man can be, he must be.'' The self-actualization level implies that anyone is capable of accomplishing anything. This also matches the concept of the top need in Maslow's hierarchy. The low levels fall in line to become the step-by-step process, and thus the basic needs must be met before achieving higher-level needs. Every person is finally capable of achieving his or her self-worth with the help of Web3 and its derivative products.

\section{Future Research Directions} \label{sec:directions}

There are many interesting and challenging directions for future research, as discussed below.

\textbf{New applications.} Within the decentralization characteristic of Web3, one of the significant transformations of application development is that people's learning and information acquisition paradigms shift from passive acceptance to proactive cognition. During the Web3 era, applications will be more open and transparent, and users will have a better understanding of the functions and goals of these Web3 products \cite{carleton2021architecting}. While DAO enables a novel governance model, it has a decisive impact on the development and usage of Web3 applications. Until now, thousands of DAOs have emerged in the financial services, investment, content creation, gaming, and social network domains. That means we are currently in the exploratory design phase of the DAO. In the foreseeable future, we will likely apply the DAO to other areas (such as healthcare, insurance, education, and urban governance \cite{kayanan2022critique}) and create revolutionary products. The booming growth of demands for virtual and real interaction in offline scenes, online virtual scenes, AR applications, digital twin space, digital collection distribution, and the design and usage of a digital human directly improves the quality of Web3 products and accelerates their iteration. Web3's developer ecosystem has created a lot of demand for new types of staff at the same time \cite{nystrom2022web3}. This will put forward higher requirements for constructing and maintaining reliable neutral platforms and practicing a decentralized concept.

\textbf{Privacy and security.} Web3's leading bug bounty platform, i.e., Immunefi\footnote{\url{https://immunefi.com}}, had reviewed massive hacking instances, as well as cases of fraud in 2022. It conducted an appalling report stating that nearly four billion dollars had been lost across the Web3 ecosystem in 2022. The report figures out that the major reason is that most users are still getting used to Web 2.0. Most people do not know how to correctly use a digital wallet to conduct transactions with others. Therefore, it is urgent to construct safer and more user-friendly software architectures \cite{carleton2021architecting, yazdinejad2020p4, yazdinejadna2021a}. In addition, some underlying technologies are also required to further improve the security services of applications, including reconciling privacy and distributed ledgers \cite{berberich2016blockchain} and cryptocurrency malware hunting \cite{yazdinejad2022crypto}. In practice, the root cause of large losses within the Web3 ecosystem is that blockchain security flaws, such as 51\% attacks, malicious nodes, and scalability issues, are not completely immune to attacks \cite{zamani2020on}.

\textbf{Power utility.} Web3 is expected to change the relations between productions and return data ownership to users. It is gratifying that Web3 technologies are gradually shifting from idealism towards pragmatism. The market for Web3 technologies has expanded dramatically over the years, and traditional Internet companies are involved too. Web3 technologies are accumulating power and building commercial value along several technology and business branches \cite{lewis2017blockchain}. However, the inefficient-by-design blockchain systems (e.g., Bitcoin) waste a lot of energy because of their proof-of-work mechanism. It is also important to consider how to make Web3 more sustainable while its technologies improve people's lives. For instance, after Ethereum completes the merger, it estimates that the update to proof of stake will cut its energy usage by 99.95\% \cite{stoll2019carbon}. Besides, a new blockchain, called Solana, combines proof of stake and proof history mechanisms to make itself faster and more efficient\footnote{Solana can deal with 65,000 transactions per second, but Ethereum is about 15 transactions per second, and Bitcoin is down to 7 transactions per second.}. This environmental issue causes the emergence of ReFi \cite{fullerton2015regenerative}. Though ReFi has been interpreted differently by institutions, it argues for tackling climate change, supporting environmental protection and biodiversity, and creating a more equitable, transparent, and sustainable financial system.

\textbf{Entertainment industry.} Entertainment activities (e.g., games \cite{thompson2021immersion}, AI-Generated Content (AIGC) \cite{roose2022ai}, interaction movies \cite{miller2020personal}, and smart tourism \cite{kontogianni2022promoting}) are almost the easiest way for encouraging users to participate in building Web3. Digital reality technology helps humans overcome the limitations of time and space on Web 2.0. For example, augmented reality and virtual reality technologies strengthen the immersive experience of users. Besides, playing games is enjoyable for all ages. As we discussed previously, Web3 has the potential to drastically transform our talents in creation, and entertainment activities will shift from being profit-driven to being interest-driven. People, especially young people, will demand and design more personalized, informative, vivid, and creative games on Web3. In return, the related technologies actually promote the development of Web3. Moreover, the training activities of teams of all sizes can benefit from these types of seamless and mixed reality interactions through the ``Playing to Learn'' concept. Games are only one of the most intuitive and addressable segments within the entertainment industry, but the Web3 opportunity extends far beyond gaming.

\section{Conclusion}  \label{sec:conclusion}

Web3 is still in its early stages, and there are still some problems and limitations with how it works that haven't been fixed yet. It encourages us to think in new ways about the World Wide Web and aims to address most problems in Web 2.0 today. This paper has presented a comprehensive overview of Web3 and its implementation. Firstly, we begin with an introduction to the motivation for decentralization and how Web3 can serve as a chance for breaking the monopoly of companies on Web 2.0. Then, we provide detailed reviews and analyses of approaches for emerging implementation challenges in Web3. The issues include data storage, communication networks, and computation. Following that, we describe various types of current Web3 applications. More importantly, we discuss the similarities and differences between Web 3.0 and Web3 in detail. A hierarchical theory of needs within Web3 is also proposed in this article. Finally, we introduce some key challenges and future research directions.

\bibliographystyle{IEEEtran}
\bibliography{paper}

\end{document}